\documentclass[runningheads]{llncs}
\usepackage[T1]{fontenc}
\usepackage{amsmath}
\usepackage{graphicx}
\usepackage{paralist}
\usepackage{microtype}
\usepackage{url}
\usepackage{float} 
\usepackage{makecell}
\usepackage{tabularx}
\usepackage{array}

\DeclareUnicodeCharacter{202F}{\,} 

\begin{document}
\title{Tokenize Everything, But Can You Sell It? RWA Liquidity Challenges and the Road Ahead}
%
%
\author{Rischan Mafrur\inst{1}\orcidID{0000-0003-4424-3736}}

\authorrunning{R. Mafrur}
%
\institute{Department of Applied Finance, Macquarie University \\
\email{rischan.mafrur@mq.edu.au}}
\maketitle              
\begin{abstract}
The tokenization of real-world assets (RWAs) promises to transform financial markets by enabling fractional ownership, global accessibility, and programmable settlement of traditionally illiquid assets such as real estate, private credit, and government bonds. While technical progress has been rapid, with over \$25 billion in tokenized RWAs brought on-chain as of 2025, liquidity remains a critical bottleneck. This paper investigates the gap between tokenization and tradability, drawing on recent academic research and market data from platforms such as RWA.xyz. We document that most RWA tokens exhibit low trading volumes, long holding periods, and limited investor participation, despite their potential for 24/7 global markets. Through case studies of tokenized real estate, private credit, and tokenized treasury funds, we present empirical liquidity observations that reveal low transfer activity, limited active address counts, and minimal secondary trading for most tokenized asset classes. Next, we categorize the structural barriers to liquidity, including regulatory gating, custodial concentration, whitelisting, valuation opacity, and lack of decentralized trading venues. Finally, we propose actionable pathways to improve liquidity, ranging from hybrid market structures and collateral-based liquidity to transparency enhancements and compliance innovation. Our findings contribute to the growing discourse on digital asset market microstructure and highlight that realizing the liquidity potential of RWAs requires coordinated progress across legal, technical, and institutional domains.

\keywords{Real World Asset  \and Liquidity Constraints \and Tokenization.}

\end{abstract}
\section{Introduction}

The tokenization of real-world assets (RWAs), the process of representing physical or financial instruments as digital tokens on a blockchain, has emerged as a prominent innovation in contemporary finance. Proponents argue that tokenization can democratize investment access and improve market efficiency by enhancing liquidity in traditionally illiquid asset classes such as real estate, private credit, and fine art~\cite{wef2022asset,arxiv2025,redstone2025}. Through fractional ownership and continuous on-chain trading, tokenization is said to lower entry barriers and facilitate broader participation in capital markets~\cite{halborn2025}.

Tokenized asset issuance has grown rapidly. As of mid-2025, over \$21 billion in tokenized RWAs (excluding stablecoins) has been deployed across more than 15 blockchain ecosystems~\cite{rwa2025,appRwa2025}. The majority of this growth is concentrated in yield generating instruments, particularly private credit, tokenized bonds, and money market funds, often backed or issued by traditional financial institutions~\cite{redstone2025}. These assets are generally integrated into decentralized finance (DeFi) protocols but are typically held for their income-generating potential rather than actively traded.

One prominent example is BlackRock’s tokenized money market fund BUIDL. Another is MakerDAO’s use of RWA tokens as collateral for issuing the DAI stablecoin. Both examples demonstrate how blockchain technology has been extended to traditional financial products. However, despite their on-chain presence, these instruments maintain relatively static liquidity profiles. Investors often pursue steady returns and avoid speculative trading. Consequently, the market has grown in size, but still follows a buy-and-hold model rather than fostering dynamic secondary market activity.

In contrast, asset classes that stand to benefit the most from improved liquidity such as real estate or fine art represent only a small portion of the RWA market. These segments, often tailored toward retail or small and medium-sized enterprises (SMEs), continue to face barriers such as regulatory complexity, limited platform access, and negligible market depth. As a result, tokenization has so far been more effective at digitizing already liquid or low volatility assets than at unlocking liquidity for structurally illiquid instruments.

This raises a central question: \textit{does tokenization, in its current implementation, genuinely improve liquidity?} While the infrastructure to tokenize nearly any asset now exists, many tokenized instruments remain difficult to sell or exit. On-chain and market data reveal persistent patterns, such as minimal trading activity, long average holding durations, and concentrated ownership~\cite{ssrn2023,kreppmeier2023}. These patterns challenge the idea that tokenization alone is sufficient to create liquid markets.

This paper investigates the liquidity profile of tokenized RWAs and explores the reasons behind the persistence of shallow markets. Drawing from both academic literature and empirical on-chain data from platforms such as RWA.xyz, we examine how investor behavior, protocol design, and regulatory limitations constrain liquidity outcomes. By comparing theoretical benefits with practical realities, we aim to identify the conditions under which tokenized RWAs might transition from static representations to actively traded financial instruments.

\section{Related Work}

\subsection{Tokenization and Liquidity: Promise vs. Reality}

A growing body of literature discusses the potential of tokenization to enhance asset liquidity by enabling 24/7 trading, fractional ownership, and streamlined settlement. Cervellati~\cite{cervellati2025} argues that tokenized markets could operate continuously, reducing frictions inherent in traditional trading systems and democratizing access to high-value, previously illiquid assets such as commercial property and fine art. Fractional ownership allows investors to participate in smaller tranches, thereby lowering entry barriers and potentially increasing market participation~\cite{cervellati2025}. Similarly, the World Economic Forum (2022) highlights fractional ownership via tokenization as a way to lower investment thresholds and increase transaction activity in traditionally illiquid asset classes~\cite{wef2022asset}.


However, academic inquiry has increasingly questioned whether these benefits are realized in practice. Swinkels~\cite{swinkels2023} conducted one of the first empirical studies on tokenized real estate markets, analyzing 58 residential property tokens issued on the RealT platform. The findings reveal a stark mismatch between theoretical and realized liquidity: on average, token ownership changed hands only once per year, suggesting extremely limited trading activity. Properties listed on decentralized exchanges like Uniswap did show approximately 25\% higher turnover than those traded peer-to-peer or via OTC channels, indicating that automated market makers (AMMs) can offer marginal improvements in liquidity~\cite{swinkels2023}. Nevertheless, turnover remained low in absolute terms and declined over time after initial issuance, highlighting the fragile and transient nature of RWA liquidity~\cite{swinkels2023}. 

Swinkels also found that legal and compliance frictions reduced tradability: investors were required to undergo off-chain whitelisting and contractual onboarding, which limited the pool of eligible secondary traders~\cite{swinkels2023}. The study concludes that fractionalization alone is insufficient to ensure liquidity; without appropriate infrastructure and legal design, tokenized assets may continue to behave like their illiquid off-chain counterparts.

A broader and more recent study by Laschinger et al.~\cite{laschinger2024} examined over 572{,}000 trades across 673 tokenized real estate properties. While the authors acknowledge the potential of decentralized markets to enhance liquidity, they also document the challenges of applying AMMs to heterogeneous, non-fungible RWAs. Unlike cryptocurrencies, each tokenized property has unique risk, legal, and valuation characteristics, which limits the efficiency of generic AMM structures. Their findings suggest that hybrid architectures, combining centralized exchanges and brokers with decentralized liquidity venues are likely necessary to facilitate robust price discovery and trading~\cite{laschinger2024}.

\subsection{Market Structure, Trust, and Regulation}

In addition to technological barriers, the literature emphasizes that investor trust and market structure are central to liquidity outcomes. An article published by The Chartered Alternative Investment Analyst Association (CAIA)~\cite{cervellati2025} highlights that confidence in the issuer, asset backing, and redemption mechanisms is essential. Without these assurances, RWA tokens may trade below their net asset value (NAV) or face wide bid ask spreads, particularly when investors are uncertain about governance or transparency. Many RWA tokens are legally classified as securities and thus restricted to accredited or KYC verified investors~\cite{swinkels2023}, which reduces the potential trader base and dampens secondary market activity. The lack of unified, standardized exchanges for security tokens further fragments liquidity, as assets are dispersed across decentralized exchanges, alternative trading systems, and private broker-dealer networks.

The International Monetary Fund (IMF)~\cite{imf2023} stresses that tokenization’s long-term impact on liquidity will depend on the development of complementary market infrastructure, including interoperable settlement systems, custody frameworks, and legal clarity on tokenized asset rights. As regulatory regimes evolve, they must strike a balance between investor protection and market fluidity.

In summary, the literature illustrates a growing consensus: while the technical infrastructure for tokenization exists, the surrounding legal, regulatory, and trust-based systems are not yet mature enough to consistently support liquid secondary markets for RWAs. This gap between aspirational and observed liquidity underlines the need for empirical monitoring, structural innovation, and thoughtful regulatory alignment~\cite{laschinger2024,cervellati2025}.

\section{Data and Methodology}

To investigate the liquidity of tokenized real-world asset (RWA) markets empirically, we draw on data from \textit{RWA.xyz}, a specialized analytics platform that aggregates on-chain metrics across public blockchains. The platform provides detailed and regularly updated data on total RWA market value, the number of issuers and holders, asset class segmentation, and protocol-level deployments. For this study, we extract summary statistics as of \textbf{July 31, 2025}, and rely on \textit{RWA.xyz}'s native asset categorization to analyze the composition and concentration of tokenized asset classes~\cite{rwa2025}.


Our methodology is twofold. First, we perform analysis of the \textit{RWA.xyz} dataset to examine the distribution of tokenized value across asset classes. This helps identify where tokenization has gained the most traction and which segments remain underrepresented. For instance, currently, the largest share of tokenized RWA value is concentrated in \textit{private credit and debt instruments}, followed by government securities such as tokenized U.S. Treasuries. In contrast, sectors like real estate and alternative collectibles represent only a small fraction of the overall tokenized asset pool. This disproportion underscores a key concern: while high-value tokenization exists, secondary market liquidity is not evenly distributed and is potentially weakest in the most illiquid asset classes.

Second, to evaluate actual liquidity, we qualitatively examining trading infrastructure and availability. For example, RWA.xyz includes metadata indicating whether specific tokens are integrated into on-chain exchange venues. The presence of AMM liquidity pools or listings on decentralized exchanges may suggest greater accessibility and continuous tradability, whereas reliance on peer-to-peer bulletin boards or off-chain brokered platforms often correlates with lower liquidity. We also consider growth in unique token holders as a proxy for potential market participation.

\section{Results and Discussion}
\subsection{Market Composition and Growth}

\begin{figure}[H]
  \centering
  \includegraphics[width=0.9\textwidth]{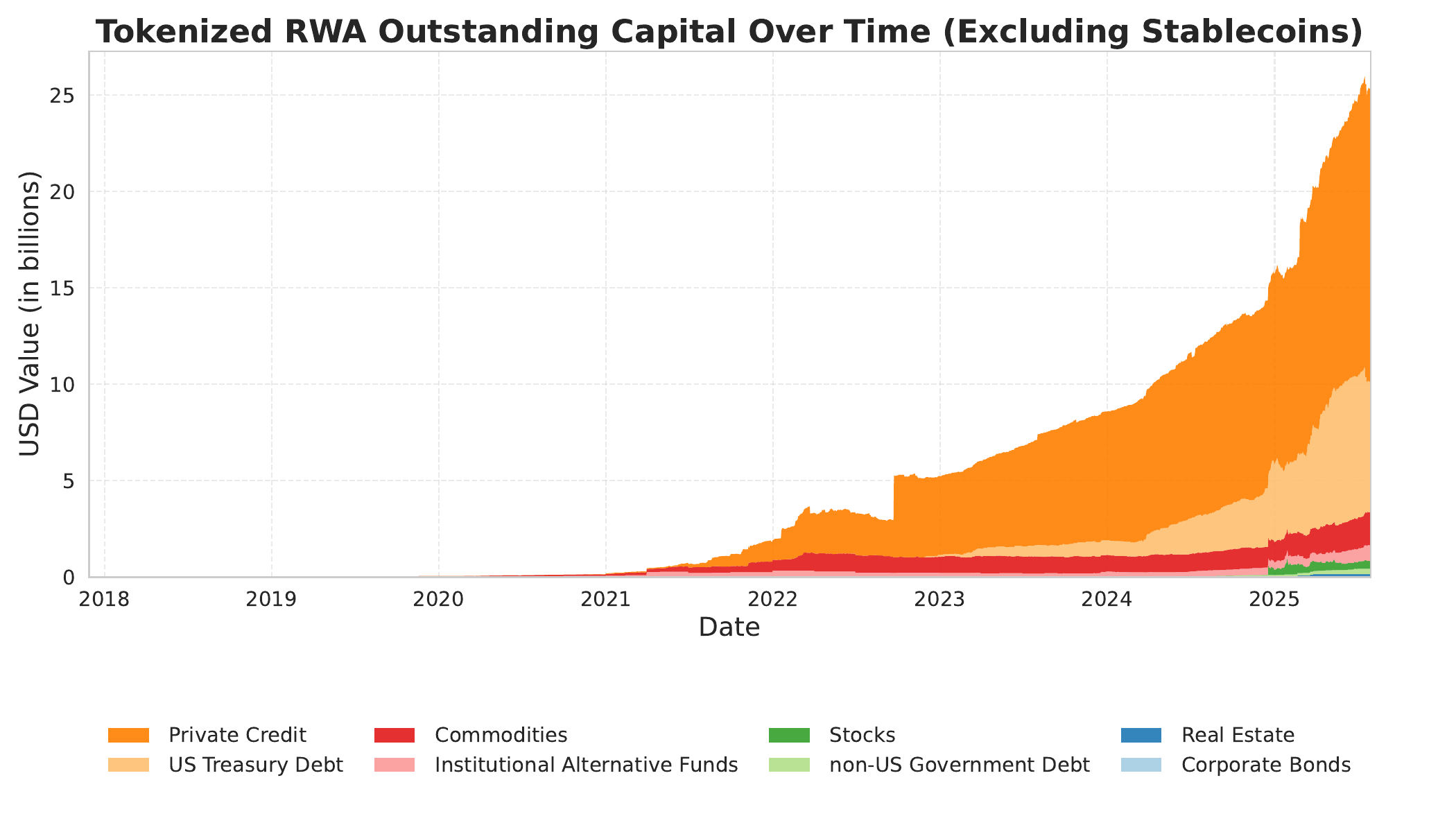}
  \caption{Tokenized RWA market growth over time.}
  \label{fig:rwa_growth}
\end{figure}

The market for tokenized real-world assets (RWAs), excluding stablecoins, has witnessed considerable growth, reaching approximately \$24--25 billion in mid-2025 as shown in Figure~\ref{fig:rwa_growth}. Table~\ref{tab:rwa-market-overview} provides a summary of the major categories, drawing on current data from RWA.xyz. Private credit and tokenized U.S. Treasuries dominate the RWA landscape, collectively accounting for the majority of non-stablecoin RWA market capitalization. This sector is driven by institutional adoption of blockchain-based debt issuance platforms, offering yield-bearing instruments that are typically held to maturity rather than actively traded. As such, liquidity remains low, and participation is often restricted to whitelisted (KYC-compliant) addresses. Other categories such as tokenized commodities (particularly gold), equities, real estate, carbon credits, trade finance instruments, and fine art remain in early stages of adoption but are gradually expanding. 

Tokenized U.S. Treasuries climbing from under \$1 billion in early 2024 to over \$7.4 billion in mid-2025. This growth is fueled by high on-chain yields and demand for secure, short-duration assets. While some secondary trading exists, these tokens generally circulate within custodial or regulated ecosystems, and remain subject to transfer restrictions under securities law. As a result, they continue to face significant liquidity challenges.

In addition, tokenized equities (approximately \$0.5 billion), real estate (approximately \$0.3 billion), and niche categories such as carbon credits, trade finance, and fine art (each around \$0.1 billion) exhibit limited liquidity and fragmented secondary markets. Many of these assets are issued under regulatory exemptions that restrict participation to accredited or whitelisted investors, thereby significantly constraining turnover. This pattern is empirically supported by Swinkels~\cite{swinkels2023}, who analyzed residential real estate tokens issued through the RealT platform. The study found extremely low turnover: on average, each token changed ownership only once per year. By comparison, equities in developed markets typically exhibit multiple turnovers annually, highlighting the stagnant nature of these tokenized markets.

\begin{table}[H]
\centering
\scriptsize
\caption{Market Overview and Trading Characteristics of Tokenized RWAs (mid-2025)}
\label{tab:rwa-market-overview}
\renewcommand\arraystretch{1.2}
\begin{tabularx}{\linewidth}{l c >{\raggedright\arraybackslash}X c c c}
\hline
\textbf{Category} & \makecell{\textbf{MCap} \\ \textbf{(USD bn)}} & \makecell{\textbf{Issuance Platforms}} & \makecell{\textbf{Secondary} \\ \textbf{Market}} & \textbf{Liquidity} & \makecell{\textbf{Transfer-} \\ \textbf{ability}} \\
\hline
Private Credit           & 14.0 & \makecell[l]{Maple, Goldfinch, \\ Figure}  & Partial & Low    & Whitelisted \\
Tokenized Treasuries     & 7.4  & \makecell[l]{Ondo, Franklin, \\ BlackRock} & Partial & Medium & Whitelisted \\
Gold (Commodities)       & 1.7  & \makecell[l]{Paxos (PAXG), \\ Tether (XAUT)} & Yes    & High   & Yes         \\
Tokenized Stocks         & 0.5  & \makecell[l]{Exodus, Swarm, \\ Backed}     & Partial & Low    & Whitelisted \\
Real Estate              & 0.3  & \makecell[l]{RealT, tZERO}              & Partial & Low    & Whitelisted \\
Carbon Credits           & 0.1  & \makecell[l]{Toucan, Moss}              & Yes     & Low    & Yes         \\
Art (Fine Art)           & 0.1  & \makecell[l]{Freeport, Maecenas}        & Partial & Low    & Whitelisted \\
\hline
\end{tabularx}
\end{table}

\subsection{Empirical Liquidity Observations}

\begin{figure}[H]
  \centering
  \includegraphics[width=0.9\textwidth]{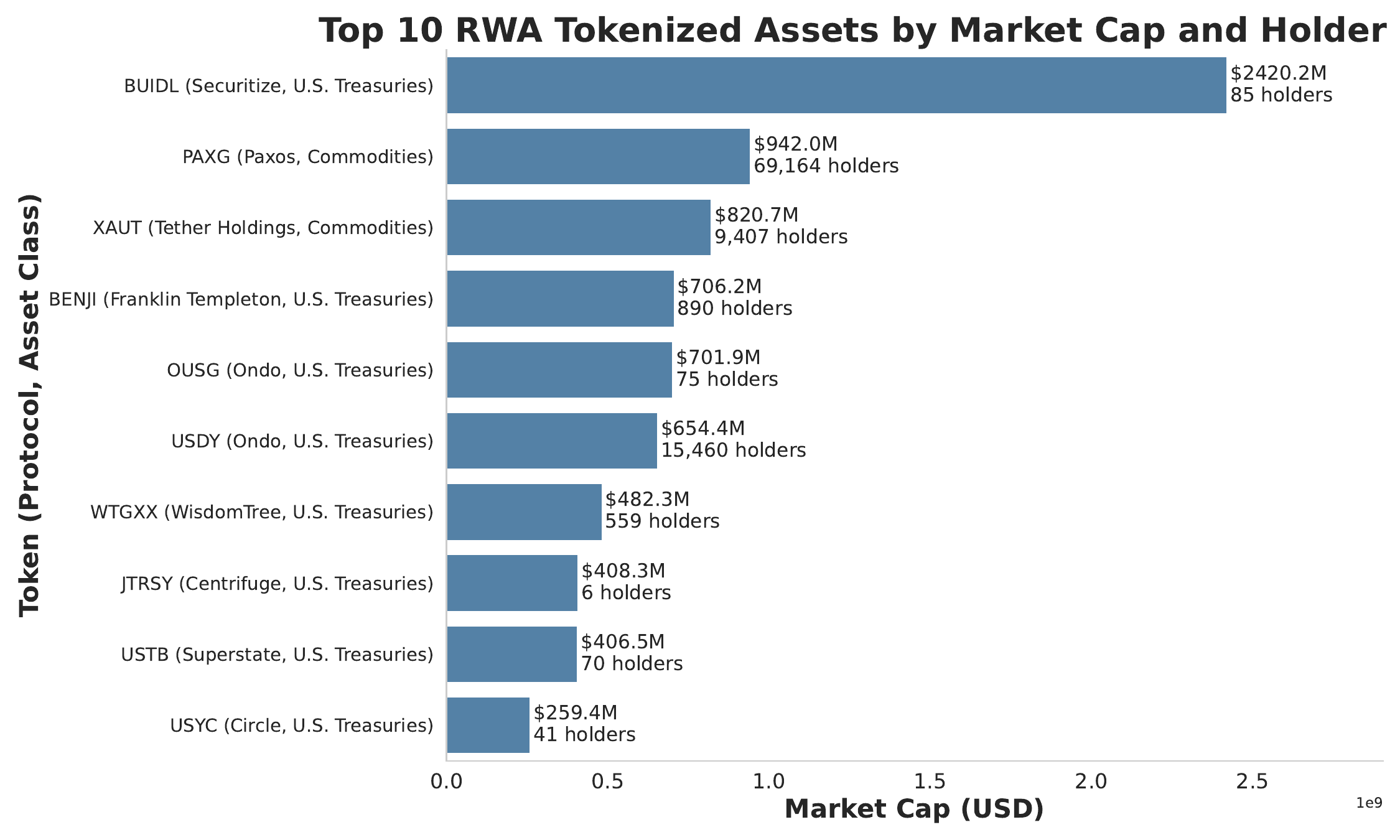}
  \caption{Top 10 tokenized RWA assets by market capitalization.}
  \label{fig:rwa_top10}
\end{figure}

\begin{table}[H]
\centering
\scriptsize
\caption{Monthly On-Chain Activity of RWA Tokens across all available blockchain networks (sourced from RWA.xyz, mid-2025)}
\label{tab:rwa_xyz_activity}
\renewcommand\arraystretch{1.2}
\begin{tabularx}{\linewidth}{l r r >{\raggedleft\arraybackslash}X >{\raggedleft\arraybackslash}X}
\hline
\textbf{Asset} & \textbf{Holders} & \makecell{\textbf{Monthly Active} \\ \textbf{Addresses}} & \makecell{\textbf{Monthly Transfer} \\ \textbf{Count}} & \makecell{\textbf{Monthly Transfer} \\ \textbf{Volume (USD)}} \\
\hline
BUIDL & 85     & 30    & 104    & 1,816,838,451 \\
PAXG  & 69,164 & 5,678 & 52,140 & 665,869,273   \\
XAUT  & 9,407  & 2,735 & 23,897 & 692,044,883   \\
OUSG  & 75     & 14    & 25     & 22,939,914    \\
BENJI & 890    & 3     & 0      & 0             \\
USDY  & 15,460 & 979   & 13,189 & 37,272,057    \\
WTGXX & 559    & 4     & 4      & 13            \\
USTB  & 70     & 14    & 704    & 78,234,900    \\
JTRSY & 6      & 2     & 2      & 381,201,262   \\
USYC  & 41     & 6     & 125    & 43,501,694    \\
\hline
\end{tabularx}
\end{table}

We base our empirical investigation on data sourced primarily from RWA.xyz and Etherscan~\cite{etherscan2025}, focusing on three key dimensions: market capitalization, token holder dispersion, and transactional activity. Figure~\ref{fig:rwa_top10} illustrates the ten largest tokenized RWA assets by market capitalization. The data reveal that the market is overwhelmingly dominated by tokenized U.S. Treasuries. However, commodity-backed tokens such as PAXG and XAUT stand out as notable exceptions, exhibiting broader holder distribution and higher levels of on-chain engagement.

This divergence can be attributed to the fact that, unlike most RWA tokens confined to permissioned or fragmented trading venues, PAXG and XAUT benefit from significantly greater liquidity. These tokens are actively traded on both centralized exchanges (e.g., Binance, Kraken) and decentralized platforms (e.g., Uniswap), facilitating broader access and enabling deeper market participation. Their presence on major exchanges supports real-time price discovery and lowers entry barriers for both retail and institutional investors. Consequently, PAXG and XAUT demonstrate higher trading volumes, wider holder dispersion, and more consistent secondary market activity relative to other RWA categories. This liquidity advantage is further reinforced by the intrinsic appeal of gold, a globally recognized store of value, which underpins the credibility and demand for these tokens.

To assess actual liquidity, we examine metrics such as monthly active addresses and transfer volumes for the top ten RWA assets by market capitalization, as shown in Table~\ref{tab:rwa_xyz_activity}. While some tokens report high market capitalizations and transactional volumes, for instance, BUIDL records over \$1.8 billion in monthly transfers, these figures often obscure underlying market thinness, with most tokens exhibiting limited user interaction and low trading frequency.

To validate and enrich these insights, we further analyze transaction-level data on the Ethereum network, the dominant blockchain for RWA token issuance by value. Table~\ref{tab:rwa_etherscan} presents on-chain activity metrics for selected RWA tokens, including total holder counts, cumulative transfer events, and temporal indicators such as \textit{active age}, \textit{unique days active}, and the \textit{longest streak} of consecutive active days. Together, these indicators provide a more granular view of token-level engagement, helping to assess the persistence and intensity of trading behavior across different asset categories.

Together, these datasets offer a comprehensive picture of the current state of tokenized RWA liquidity. Four main observations emerge:

\paragraph{1. Market Capitalization vs. Trading Volume.}  
While private credit and U.S. Treasuries collectively exceed \$20 billion in tokenized value, on-chain transfer activity remains sparse. Treasury tokens are primarily transacted during mint and redemption events, with little ongoing trading. In contrast, commodity-backed tokens like PAXG demonstrate a more active profile, with \$665 million in monthly transfer volume (Table~\ref{tab:rwa_xyz_activity}) and over 1.4 million historical transfers on Ethereum alone (Table~\ref{tab:rwa_etherscan}). This divergence highlights a mismatch between token issuance and actual market liquidity.

\paragraph{2. Token Holdership vs. Transfer Activity.}  
Table~\ref{fig:rwa_top10} shows that most RWA tokens are seldom traded. For instance, BUIDL, which is the RWA asset with the highest market capitalization, has only 85 holders, 30 monthly active addresses, and 104 monthly transfers. Conversely, PAXG supports over 52,000 transfers per month among its ~69,000 holders, reflecting a modest but sustained level of secondary market activity. The majority of RWAs exhibit minimal transfer velocity, suggesting passive, long-term holding behavior rather than active trading. In addition to low transfer volumes, streak-based metrics reveal a lack of consistent trading activity. Most RWA tokens exhibit short “longest streak” durations, for instance, JTRSY (3 days) and WTGXX (4 days) indicating that transfers happen in isolated bursts rather than sustained trading cycles. In contrast, PAXG maintains over five years of uninterrupted activity, illustrating the deep liquidity that is largely absent in other RWA markets (Table~\ref{tab:rwa_etherscan}).

\paragraph{3. Stagnant Asset Flows and Absence of Secondary Markets.}  
Many RWA tokens lack access to open exchanges and instead operate under permissioned or whitelist-only systems. Tokens like ACRED are restricted to accredited investors and enforce transfer controls at the smart contract level. As a result, price formation is often tied to administrative NAV updates rather than market-driven discovery. In extreme cases, such as reinsurance fund tokens, holdings are concentrated in just one or two addresses, with no observed transfer activity for months.

\paragraph{4. Platforms Built for Primary Issuance, Not Trading.}  
Most RWA protocols prioritize issuance and asset onboarding over secondary liquidity. Maple, for example, facilitates tokenized loan origination but lacks integrated exit mechanisms, requiring bilateral negotiation for token transfers. Similarly, RealT restricts trading to platform-managed OTC channels with stringent KYC requirements. As shown in Table~\ref{tab:rwa_xyz_activity}, many institutional-grade tokens have fewer than ten active addresses per month, reinforcing the conclusion that these systems are not structured for fluid secondary market activity.

These findings underscore a persistent gap between the theoretical liquidity promised by tokenization and the empirical reality of usage on-chain. While the infrastructure for issuance is advancing rapidly, secondary market development remains constrained by regulatory design, user access barriers, and the absence of trading incentives.

\begin{table}[H]
\centering
\scriptsize
\caption{On-Chain Activity of Selected RWA Tokens on Ethereum (via Etherscan)}
\label{tab:rwa_etherscan}
\renewcommand\arraystretch{1.2}
\begin{tabularx}{\linewidth}{l r r X X X}
\hline
\textbf{Asset} & \textbf{Holders} & \textbf{Transfers} & \makecell{\textbf{Active Age} \\ \textbf{(days)}} & \makecell{\textbf{Unique Days} \\ \textbf{Active}} & \makecell{\textbf{Longest} \\ \textbf{Streak}} \\
\hline
BUIDL   & 49     & 6,977     & 1 Year 151 Days    & 1 Year 10 Days      & 61 Days       \\
PAXG    & 50,140 & 1,427,599 & 5 Years 339 Days   & 5 Years 331 Days    & 5 Years 304 Days \\
XAUt    & 8,935  & 188,153   & 3 Years 267 Days   & 3 Years 260 Days    & 2 Years 86 Days  \\
OUSG    & 56     & 1,546     & 2 Years 188 Days   & 1 Year 182 Days     & 13 Days       \\
BENJI   & 3      & 573       & 255 Days           & 190 Days            & 62 Days       \\
USDY    & 372    & 3,343     & 1 Year 319 Days    & 1 Year 140 Days     & 48 Days       \\
WTGXX   & 11     & 132       & 320 Days           & 62 Days             & 4 Days        \\
USTB    & 53     & 2,773     & 1 Year 211 Days    & 303 Days            & 56 Days       \\
JTRSY   & 4      & 34        & 360 Days           & 20 Days             & 3 Days        \\
USYC    & 33     & 37,176    & 2 Years 50 Days    & 1 Year 293 Days     & 250 Days      \\
\hline
\end{tabularx}
\end{table}

\subsection{Structural Barriers to Liquidity}

Several structural factors reinforce this illiquidity:

\begin{enumerate}

  \item \textbf{Fragmented Marketplaces:} Unlike traditional equity markets where most trades occur on centralized exchanges like NYSE or NASDAQ, RWA markets are fragmented across decentralized platforms (e.g., Uniswap), specialized custodial systems (e.g., tZERO), and informal OTC channels. This lack of centralized trading venues hinders liquidity aggregation and price discovery.

  \item \textbf{Regulatory Restrictions:} Many tokens are issued under frameworks that limit access to accredited or KYC-verified investors. Jurisdictional limitations, onboarding frictions, and lock-up periods reduce the number of active participants, thereby lowering turnover and increasing time-to-trade~\cite{swinkels2023}.

  \item \textbf{Valuation Uncertainty:} Illiquidity is compounded by difficulty in establishing fair value. Traders may be uncertain how to price tokens tied to unique or opaque assets, such as individual real estate holdings or private loans. As a result, bid-ask spreads widen, and investors often apply a liquidity discount~\cite{cervellati2025}, creating a self-reinforcing illiquidity spiral.

  \item \textbf{Lack of Market Makers:} Traditional financial markets rely on dedicated market makers to ensure liquidity and narrow spreads. In the tokenized RWA space, such actors are rare. While some DeFi protocols incentivize liquidity providers, these incentives are often insufficient or inefficient when applied to non-fungible, low-volume tokens~\cite{laschinger2024}.

  \item \textbf{Technological and Operational issues:} Beyond wallet management and protocol complexity, users face high gas fees, limited scalability, and interoperability issues across blockchain networks. While blockchains enable decentralized asset settlement, they also introduce latency and cost barriers particularly on networks like Ethereum where congestion leads to elevated transaction fees. Moreover, RWAs issued on siloed chains (e.g., private or permissioned blockchains) lack cross-chain compatibility, restricting composability with DeFi protocols and fragmenting liquidity even further. Until seamless multi-chain infrastructure and lower-cost execution layers become standard, these frictions will continue to impede RWA market development.

\end{enumerate}

\subsection{Pathways to Improved Liquidity}

\begin{table}[H]
\scriptsize
\centering
\caption{Regulatory, Risk, and Integration Characteristics of Tokenized RWAs}
\renewcommand\arraystretch{1.2}
\begin{tabularx}{\linewidth}{l c c c >{\raggedright\arraybackslash}X}
\hline
\textbf{Category} & \makecell{\textbf{Regulatory} \\ \textbf{Classification}} & \textbf{Redemption} & \makecell{\textbf{Custodial} \\ \textbf{Risk}} & \makecell{\textbf{On-Chain Integration}} \\
\hline
Private Credit      & Security   & No       & Medium & \makecell[l]{Centrifuge (via MakerDAO); \\ Maple Finance; Goldfinch} \\
Tokenized Treasuries & Security  & Yes      & Low    & \makecell[l]{MakerDAO; Ondo Finance; \\ Maple Finance} \\
Gold (Commodities)  & Commodity  & Yes      & Medium & \makecell[l]{Aave; MakerDAO (PAXG)} \\
Tokenized Stocks    & Security   & No       & Medium & None \\
Real Estate         & Security   & No       & High   & None \\
Carbon Credits      & Commodity  & No       & High   & \makecell[l]{KlimaDAO; Flowcarbon \\ (Centrifuge)} \\
Art (Fine Art)      & Security   & No       & Medium & None \\
\hline
\end{tabularx}
\label{tab:regulatory-overview}
\end{table}

Our findings suggest that liquidity constraints in tokenized real-world asset (RWA) markets are systemic rather than incidental. Table~\ref{tab:regulatory-overview} highlights key frictions: the dominance of security classifications, limited redemption options, custodial dependencies, and uneven on-chain integration. Addressing these bottlenecks requires a multidimensional approach.

\begin{enumerate}
    \item \textbf{Hybrid Market Structures:} As shown in Table~\ref{tab:regulatory-overview}, the majority of tokenized RWAs are classified as securities and exhibit limited on-chain integration. This is particularly evident in sectors such as tokenized stocks, real estate, and fine art, where integration is often listed as “None.” The absence of seamless connectivity to decentralized finance (DeFi) infrastructure restricts key functionalities such as programmability, composability, and automated market-making. A hybrid model that combines regulated, centralized platforms for primary issuance and compliance with decentralized protocols for secondary trading could help unlock liquidity, particularly in illiquid sectors like tokenized real estate and alternative assets.

    \item \textbf{Incentives for Liquidity Providers:} Low observed liquidity in categories such as tokenized stocks (Table~\ref{tab:rwa-market-overview}) reflects the absence of active market-making. Protocols could implement structured incentives (e.g., allocating a portion of bond yields or protocol fees to liquidity providers). This is particularly pertinent for whitelisted assets that suffer from limited secondary market traction. 

    \item \textbf{Improved Transparency and Standardized Valuation:} One recurring challenge across sectors is valuation opacity. Private credit and real estate tokens often lack continuous pricing benchmarks, creating bid-ask asymmetry and deterring trade. Improved disclosures, on-chain performance metrics, and third-party appraisals especially for high custody risk sectors like real estate and carbon credits can narrow price uncertainty. As CAIA notes, reducing information asymmetry reduces investor risk premiums and improves market functioning~\cite{cervellati2025}.

    \item \textbf{Regulatory Modernization and Broader Access:} As Table~\ref{tab:regulatory-overview} shows, most RWAs are issued as restricted securities, accessible only to KYC-verified or accredited investors. Expanding access through regulatory innovations (e.g., Reg A+, the EU Pilot Regime) and leveraging blockchain-native compliance tooling (e.g., automated KYC/AML, permissioned tokens) could broaden the eligible investor base. This is especially impactful in markets like tokenized treasuries which already attract institutional interest but lack widespread retail access.

    \item \textbf{Liquidity Through Collateralization and Lending:} Not all liquidity must come via direct trading. Collateralization protocols allow investors to access liquidity without selling their assets. For instance, MakerDAO has recently begun accepting tokenized U.S. Treasury instruments such as short‑term T‑bills as collateral for borrowing DAI, following its Spark Tokenization Grand Prix and large-scale reserve allocations. To date, private credit tokens do not appear to be officially supported collateral types~\cite{ranglani2022rwa}. This form of indirect liquidity is especially important for long-duration or low-turnover assets (e.g., real estate or fine art). By reducing forced sales and unlocking capital efficiency, such mechanisms serve as a release valve for otherwise illiquid holdings.

    \item \textbf{Institutional-Grade Custody and Secondary Markets:} As seen in Table~\ref{tab:regulatory-overview}, categories with high custodial risk (e.g., real estate, carbon credits) often exhibit the lowest liquidity. Institutional-grade custody solutions and emerging regulated token exchanges (e.g., EU DLT MTFs) can provide the infrastructure needed to foster secure, scalable, and compliant trading venues. These platforms are increasingly supported by traditional financial institutions, enhancing trust and reducing execution barriers.

    \item \textbf{Decentralization (on-chain) as a Liquidity Catalyst:} While much of today’s RWA infrastructure operates under centralized issuance and custody, decentralization still holds potential to improve liquidity and reduce systemic risks if applied meaningfully. Permissionless trading venues, DAO-based governance, and composable DeFi protocols can reduce custodial dependencies and expand access. Decentralized automated market makers (AMMs), for example, enable continuous, global trading of RWA tokens without relying on centralized intermediaries, thereby enhancing price discovery and reducing barriers to entry. Recent studies on tokenized ecosystems warn, however, that decentralization is often more nominal than actual: control over infrastructure, governance, or off-chain data frequently remains centralized~\cite{mafrur2025}. For decentralization to meaningfully impact RWA liquidity, it must extend beyond protocol branding to encompass transparent compliance primitives, open liquidity rails, and distributed operational control particularly in sectors where reliance on centralized actors limits transferability and exit options.

    \item \textbf{Market Transparency and Audience Education:}  
    A key but often overlooked factor in improving RWA liquidity is the need for broader audience education and accessible analytics infrastructure. At present, comprehensive data platforms for RWAs remain limited. RWA.xyz is the primary public-facing dashboard for on-chain RWA metrics, whereas the broader crypto ecosystem benefits from multiple well-established data aggregators such as CoinMarketCap, CoinGecko, Messari, etc. This disparity results in limited visibility, fragmented price discovery, and insufficient investor confidence. Expanding the availability of analytics tools, dashboards, and explanatory content (e.g., blogs, research platforms, and whitepapers) tailored to both institutional and retail users can enhance transparency and demystify asset behavior. In parallel, increased academic engagement, including empirical studies, market microstructure research, and standardized reporting frameworks can bridge knowledge gaps and support more informed participation. As work in~\cite{mafrur2025blockchain} argues in the context of blockchain data analytics, advancing tooling and educational infrastructure is critical for reducing market opacity and aligning academic and industry progress. By fostering better-informed market participants, educational initiatives can indirectly support more vibrant secondary markets and reduce liquidity risk.

\end{enumerate}

In summary, liquidity improvement is not a singular fix but a layered architecture of legal, technical, and market interventions. By addressing regulatory access, market structure inefficiencies, valuation uncertainty, and the role of on-chain integration, the RWA ecosystem can transition from issuance-centric to transaction-centric design. It is important to note that not all RWAs require high-frequency liquidity. Some instruments such as closed-end funds or bespoke debt agreements trade infrequently even in traditional finance. However, ensuring optionality in exit and reducing the liquidity discount remain vital goals.

\section{Conclusion}

\textit{“Yes, we can tokenize everything, but can we trade it?”} This paper set out to explore a critical tension in the evolving landscape of blockchain-based finance: while nearly any asset can be technically represented as a token on-chain, ensuring those tokens are actively tradable remains a far more difficult challenge.

Drawing from both aggregate market data and detailed empirical case studies, we find that liquidity is the enduring bottleneck in the promise of real-world asset (RWA) tokenization. The technological capabilities have advanced rapidly, enabling a wide variety of assets ranging from U.S. Treasuries to fractionalized real estate to be tokenized and issued across multiple blockchain networks. Yet the majority of these tokens, particularly those tied to traditionally illiquid assets, exhibit minimal trading activity, wide bid–ask spreads, and structural barriers to exit.

Crucially, this illiquidity is not caused by tokenization itself, but by underdeveloped market infrastructure, fragmented trading venues, restrictive regulations, and cautious investor sentiment. In many cases, assets have been successfully tokenized, but the ability to trade them efficiently has not kept pace. In other words, we have digitized ownership, but have not yet built the supporting systems needed to make these digital assets liquid and accessible in practice.

Nonetheless, the future of tokenized finance remains promising provided that the liquidity problem is addressed directly. This includes regulatory modernization to enable broader participation, technological solutions such as hybrid market structures and cross-chain interoperability, and financial engineering to incentivize liquidity provision. Emerging models are already experimenting with alternative forms of liquidity. For example, MakerDAO’s integration of RWAs into its stablecoin collateral system allows users to borrow DAI against tokenized real-world assets, effectively realizing liquidity without immediate asset sales. Similarly, the rise of regulated token exchanges many backed by institutional players suggests that trustworthy, scalable secondary markets are in development.

In short, tokenization has shown that it works technically and has seen growing adoption across various asset classes. However, without reliable and efficient ways to trade these tokens, their impact on financial markets will remain limited. Improving liquidity should not be treated as a byproduct of market growth, but as a core design goal. Practical steps such as better on-chain integration, interoperable platforms, and clear regulatory frameworks are needed to make tokenized assets more usable and accessible. Only then can tokenization deliver real benefits beyond digitization, such as broader access and faster settlement.

\bibliographystyle{splncs04} 
\bibliography{rwa_paper}            

\end{document}